\documentclass{article}
\usepackage{authblk}
\usepackage{geometry}
\usepackage{graphicx}
\usepackage{caption}
\usepackage{xcolor}
\usepackage{float}
\usepackage{soul}
\graphicspath{Images}
\usepackage{amsmath, amssymb}
\usepackage{hyperref}
\usepackage{cancel}
\usepackage[fleqn]{mathtools}

\usepackage{cancel}
\usepackage{chngcntr}
\counterwithin*{equation}{section}
\counterwithin*{equation}{subsection}

\usepackage [english]{babel}
\usepackage{relsize}
\usepackage{multirow}
\usepackage{bigints}
\usepackage{amsmath, amssymb, physics, bm}
\usepackage{hyperref}
\usepackage{amsthm}
\theoremstyle{remark}

\usepackage{tikz}
\usetikzlibrary{arrows.meta,calc}
\usepackage[toc,page]{appendix}

\usepackage{authblk}
\usepackage{orcidlink}
\usepackage{hyperref}
\usepackage{tabularx}
\usepackage{booktabs}
\usepackage{threeparttable}

\title{A regularisation method to obtain analytical solutions to the de Broglie--Bohm wave equation}

\author{Anand Aruna Kumar$^{1,*,\dagger}$\orcidlink{0000-0001-6148-2777}, S.K. Srivatsa $^{2,\dagger}$\orcidlink{0000-0002-4808-9270} and Rajesh Tengli$^{3,\dagger}$\orcidlink{0000-0003-2433-1245}}

\affil[1]{Hardware Research Engineer, IBM Research, Albany,NY,  USA}

\affil[2]{Director, Numera.AI Labs Pvt. Ltd., Bangalore, India}

\affil[3]{Consultant, Volvo, Sweden}

\affil[$*$]{Author for correspondence: \href{mailto:anand.aruna.kumar@ibm.com}{anand.aruna.kumar@ibm.com}}
\affil{$^\dagger$Private Research}

\date{} 

\begin{document}
	\maketitle
	
	\begin{abstract}

		We develop a variational regularisation framework that enables analytical solutions of the stationary de Broglie--Bohm  wave equation. The formulation begins with a Fisher-information--augmented action functional for the probability density and phase fields, yielding the Madelung (Hamilton--Jacobi and continuity) equations and, upon complex recombination, a Schr\"{o}dinger-type equation with a parametric information coupling $\mu$. 
		\vspace{0.5pc}
		
        Beyond the density-based formulation, we develop a variational regularisation scheme for the de Broglie-–Bohm equations, introducing both a global Fisher-information-based regularisation at the level of the action functional and a shell-level regularisation emerging from stationary flux closure, which together enable exact analytical reductions of the Madelung system. At the shell level, this reduction isolates the regularisation mechanism in the spatial momentum flow and yields constrained Euler--Lagrange equations governing admissible amplitude configurations. The resulting first integral possesses an elliptic (Weierstrass) structure whose admissible asymptotic branch enforces a universal canonical relation $p(x)x \to \mu/2$ near amplitude zeros. This canonical product arises dynamically from amplitude regularity and conserved flux, rather  than being postulated, and reflects the consistent action of regularisation at both the global variational and local shell levels.
		\vspace{0.5pc}
		
		The framework yields closed-form analytical solutions for standard potentials and reveals a systematic inverse-square regularising term in the effective potential. The associated elliptic discriminant defines a geometric length scale which, under physical identification $\mu=\hbar$, naturally reduces to the reduced Compton wavelength. Thus, canonical Bohmian regularisation is interpreted as a variational admissibility condition on density dynamics, providing structurally stable analytical branches and modified yet consistent energy spectra within stationary dBB mechanics.
		
		\vspace{0.5pc}
	
\end{abstract}

	
\section{Introduction}

The de~Broglie--Bohm (dBB) formulation of quantum mechanics is guided by a real-valued
pilot-wave amplitude and phase field obtained from the polar decomposition
$\psi = X e^{iS/\mu}$, where $X=\sqrt{P}$ is the amplitude and
$P=|\psi|^2$ denotes the probability density, while $S$ is the
Hamilton--Jacobi (HJ) phase. Substitution of this polar form into the
Schr\"{o}dinger equation yields coupled equations: a modified HJ equation
containing the quantum potential and a continuity equation enforcing
probability conservation. Analytical solutions of these coupled nonlinear
equations remain available only for a limited set of potentials, motivating
the development of systematic analytical reduction and regularisation
schemes.
\vspace{0.5pc}

\noindent Several studies in the dBB framework have emphasised analytical or semi-analytical
solutions of the wave function and associated trajectories for specific quantum
systems \cite{PeterH}, including applications to quantum cosmology \cite{nelsonbrogliebohm}, singular wave functions,
and trajectory-based formulations of quantum dynamics~\cite{durr2023quantum,
sanz2020trajectory,passon2019debrogliebohm,sanz2024bohmian}. These approaches highlight both the usefulness and the
difficulty of obtaining closed-form solutions of the Bohmian equations, particularly
due to the nonlinear quantum potential term. Consequently, identifying structural
principles that enable exact analytical reductions remains an important open problem
within the dBB programme.
\vspace{0.5pc}

\noindent Related variational and Hamiltonian treatments of the Madelung and
dBB systems have been discussed in the literature, where the
hydrodynamic formulation is embedded within a HJ framework and
extended to include particle back-reaction and Liouvillian structures
\cite{PeterHNUC1,PeterHNUC2}. From an information-theoretic viewpoint,
Reginatto \cite{Reginatto1998} has shown that supplementing classical HJ dynamics with
a minimum Fisher information constraint leads to the Schr\"{o}dinger equation in
hydrodynamic variables, thereby providing a variational bridge between classical
and quantum dynamics. Analytical approaches to the Madelung equations have also
been explored using scale-invariant functional ans\"atze, notably by
Barna \textit{et al.}~\cite{AnsolMadelung}, demonstrating the feasibility of
closed-form solutions under suitable structural assumptions.
\vspace{0.5pc}

\noindent In this context, the present work extends the variational perspective by
introducing a stationary flux closure together with a constraint-based
regularisation that enables explicit analytical reductions. By augmenting the
classical HJ action with a Fisher information \cite{Fri} gradient penalty,
one obtains a density functional whose Euler--Lagrange (EL) equations reproduce the
dBB system and, upon complex recombination, a Schr\"{o}dinger-type equation with a
parametric coupling constant~$\mu$. This construction interpolates between
classical mechanics ($\mu\to0$) and standard quantum mechanics ($\mu=\hbar$),
and leads, in the stationary regime, to Sturm--Liouville structures and
Ermakov--Lewis \cite{Ermakov1880} invariant formulations that permit exact analytical solution
branches.
\vspace{0.5pc}

\noindent However, the density-based formulation alone does not uniquely determine
the admissible spatial behaviour of the local momentum field $p=\nabla S$.
In stationary flows, the continuity equation enforces a conserved flux relation,
\(
\nabla\!\cdot(P\,\mathbf{p})=0,
\)
which in one dimension reduces to a constant current condition $P(x)p(x)=C$.
This stationary flux closure ties the momentum field pointwise to the amplitude but
still leaves a residual freedom in admissible momentum profiles compatible
with normalisable densities. Controlling this freedom requires a local
regularisation principle acting directly on the momentum sector and consistent
with the HJ energy balance in the effective potential $V_{\mathrm{eff}}=V+Q$.
\vspace{0.5pc}

\noindent In the present work we formulate such a principle through a global variational framework and its reduced (shell)
variational description obtained after imposing the stationary flux closure,
i.e.\ the steady-state condition $\nabla\!\cdot(P\mathbf{p})=0$ leading to
$P(x)p(x)=C$. The resulting constrained action depends only on the amplitude field, subject to a normalisation constraint, with the momentum field determined pointwise through the flux relation.
The corresponding EL equation yields a first integral whose structure is
generically elliptic and admits a Weierstrass representation. Analysis of the
admissible asymptotic branch near amplitude zeros reveals a universal canonical relation
\[
p(x)\,x \to \frac{\mu}{2},
\]
which emerges dynamically from amplitude regularity and conserved flux rather  than
being imposed as an external constraint.
\vspace{0.5pc}

\noindent This canonical Bohmian regularisation introduces an inverse-square contribution in
the effective potential, producing analytically tractable wave functions and slight
spectral shifts while preserving the overall spectral structure. Furthermore, the
elliptic discriminant defines a characteristic geometric length scale which, under
the physical identification $\mu=\hbar$ and $E\sim mc^2$, reduces naturally to the
reduced Compton wavelength. This suggests that short-distance limitations arise as a
geometric consequence of variational admissibility of the density dynamics rather
than as an additional interpretive postulate.
\vspace{0.5pc}

\noindent The present work therefore develops a unified analytical framework for
stationary dBB systems based on three complementary variational reductions:\\
\noindent (i) a Fisher information density functional leading to a Schr\"{o}dinger-type
amplitude equation and globally constrained spectral structure,\\
\noindent (ii) a global amplitude admissibility principle, yielding an Ermakov--Pinney
reduction that encodes regularity and finite trajectory behaviour, and\\
\noindent (iii) a reduced shell variational formulation that isolates the local
regularisation mechanism and selects physically admissible canonical branches.

\section{Framework for wave function in polar form}
\label{sec2}

\noindent In this section, we derive the polar-form (`ansatz') wave function, by coupling classical HJ and continuity equations with information theory's error quantification term. 

\subsection{Combined action for HJ and continuity equations}
\
\noindent The classical HJ and continuity equations for a single particle are respectively:
\begin{subequations}
	\begin{equation}
		\label{eq:clhj}
		\partial_t S + \frac{(\nabla S)^2}{2m} + V(r,t)=0
	\end{equation}
	and
	\begin{equation}
		\label{eq:clcon}
		\partial_t P + \nabla\!\cdot(P \textbf{v}\big)=0, \text{  where  } \textbf{v}=\frac{\nabla S}{m}.
	\end{equation}
\end{subequations}
\noindent Here, $P$ is an unknown probability distribution function (PDF) that is to be determined, and is unique to the potential.
\vspace{0.5pc}
\noindent With a bit of algebra, the above two equations can be combined into one equation

\begin{subequations}
	\begin{equation}
		\partial_t(PS) + \nabla\cdot\big(S P\mathbf v\big) +\; P\Big( V - \frac{(\nabla S)^2}{2m}\Big)=0
	\end{equation}
	\noindent Isolating divergence term on the left, one can express it as shown below
	\begin{equation}
		\boxed{\partial_t(PS) + \nabla\cdot\big(S P\mathbf v\big) =P\left(\frac{(\nabla S)^2}{2m} -V\right) }
	\end{equation}
\end{subequations}

\noindent The left-hand side, $\partial_t(PS) + \nabla\!\cdot(SP\textbf{v})$, represents the
transport (conservation) form of the density-weighted phase $P S$ advected
with velocity $\textbf{v}=\nabla S/m$. The right-hand side acts as a local source term
that depends on the balance between kinetic and potential energy densities.
This structure naturally suggests the construction of a variational functional
with fields $P$ and $S$.
\vspace{0.5pc}

\noindent A suitable action for the above PDEs can be constructed with variables as $P$ and $S$, and can be written as \cite{Gitkup}.
\begin{equation}
	\label{eq:classact}
	\mathcal{A}_0[P,S] \;=\; \int dt\!\int d^3r\; P\Big(\partial_t S + \frac{(\nabla S)^2}{2m}+V\Big)
\end{equation}

\noindent and extremisation of this action with respect to $P$ {$\rightarrow$} $\delta_P\mathcal A_0$  gives us the (\ref{eq:clhj}) and  with respect to $S$  {$\rightarrow$}  $\delta_S\mathcal A_0$ gives (\ref{eq:clcon}).
\vspace{0.5pc}

\noindent The functional $\mathcal A_0[P,S]$ should be regarded as a constrained
\emph{stationary} functional rather than a time-dependent action.
Accordingly, its integrand has the structure of a spatial energy density
written in the non-canonical field variables $(P,S)$, with contributions from
the kinetic and potential sectors together with an internal gradient term.
The appearance of a $T+V$ -type structure, rather than the canonical $T-V$
form, reflects this non-canonical parametrisation and the fact that no
Legendre transformation (to canonical variables) has been performed.
\subsection{Revised action with Fisher information term}
\vspace{0.5pc}

\noindent The derived action, (\ref{eq:classact}) is self-consistent for solving the classical mechanical system of a single particle. The PDF is an unknown quantity that must be determined based on the potential governing a specific type of motion. Physical parameters and other variables must be measured and compared with the model. This comparison, in turn, allows us to determine the PDF to the required accuracy. Since any measurement process naturally yields a data distribution, a statistical interpretation is needed, as the data will inevitably have variance due to measurement errors. Therefore, a quantifiable error function is necessary for a complete description. 
\vspace{0.5pc}

\noindent Information entropy is one such  measure of uncertainty in a measurable system. The PDF can be obtained by performing a series of measurements on the particle’s motion. In contrast to macroscopic classical systems, whose trajectories can be deterministically tracked, microscopic systems—even single particles—are
usually characterised operationally through repeated measurements, due to the
practical difficulty of directly determining their instantaneous dynamical
quantities. Consequently, errors are an unavoidable part of the measurement process and must be rigorously quantified. A scientific method for quantifying the minimum possible error is achieved by utilising the Fisher information ($I$). The Fisher information measures the amount of information an observable carries about an unknown parameter (like the system's position or momentum) and is formally expressed as:
\begin{equation}
	I(\theta) = E\left[ \left(\frac{\partial}{\partial \theta} \ln w(y|\theta) \right)^2 \right] = \int \left(\frac{\partial}{\partial \theta} \ln w(y|\theta) \right)^2 \cdot \, w(y|\theta)  dy
\end{equation}
where $w(y|\theta)$ is the likelihood function (PDF) that depends on the parameter $\theta$. The Fisher information is crucial because it sets a fundamental limit on the achievable precision. Specifically,  the Fisher information ($I$) and the variance of an unbiased estimator, $\mathrm{Var}(\hat{\theta})$, are related by the Cramér-Rao Inequality.  If we denote the minimum standard error (or standard deviation) of the estimator as $e$, then the variance is $e^2$, and the inequality can be written as \cite{Cramer1946}:
\begin{equation}
	\mathrm{Var}(\hat{\theta}) = e^2 \ge \frac{1}{I(\theta)}  \text{ or }e^2I(\theta) \geq 1 
\end{equation}
This inequality establishes that the reciprocal of the Fisher information is the lowest possible variance (or error) one can achieve in determining the physical quantity, providing a complete description of the error inherent in the system. 
\vspace{0.5pc}

\noindent A generalisation of the same expression for space and time dependent variables is given by
\begin{equation}
	{I}[P] =\int dt\int d^3r\frac{(\nabla P)^2}{P} = {4}\int dt\int d^3r\; (\nabla \sqrt{P})^2
\end{equation}
where $P$ is the PDF, now a density term.  We couple this relation to the HJ and continuity equation to obtain statistical description of the system. It is now possible to write a Lagrangian with information coupling, $I$. A suitable dimension factor, $\mu^2/8m$ is introduced to ensure that the Fisher information has the correct dimension of energy density. \\
\vspace{0.5pc} 

\noindent The introduction of a Fisher information gradient term in the action functional
is motivated by broader information-theoretic variational principles. In
particular, Frieden and Gatenby showed that extending Hardy's axioms of
physics to statistical systems leads to a principle of maximum Fisher
information, \(I=I_{\max}\), from which several physical laws, including
quantum wave equations, may be derived within the Extreme Physical Information
(EPI) framework~\cite{FriGat}. In this spirit, the Fisher information gradient
penalty employed below may be viewed as a constrained information functional
that couples the amplitude gradients to the classical Hamilton--Jacobi action.
\vspace{0.5pc}

\noindent The error functional is
\begin{equation}
	\mathcal{I}[P] =\frac{\mu^2}{8m}\!\int dt\!\int d^3r \frac{(\nabla P)^2}{P} =\frac{\mu^2}{2m}\!\int dt\!\int d^3r \; {|\nabla\sqrt{P}|^2}
\end{equation}
\noindent  Adding this term with an appropriate sign to $\mathcal A_0$ , we have the canonical Lagrangian density.
\begin{equation}
	\label{eq:density_action}
	{\mathcal A[P,S] = \int dt\int d^3r\left(P\left(\partial_t S + \frac{(\nabla S)^2}{2m}+V\right) - \frac{\mu^2}{2m}|\nabla\sqrt{P}|^2\right)}
\end{equation}
\noindent With this modified action, the variation $\mathcal \delta A/\delta S$ produces the continuity equation
\begin{subequations}
	\begin{equation}
		\partial_t P + \nabla\!\cdot(P \textbf{v}\big)=0
	\end{equation}
	and  $\delta A/\delta P$ results in Madelung or dBB-like equation:
	\begin{equation}
		\partial_t S + \frac{(\nabla S)^2}{2m}+V -\frac{\mu^2}{2m}\frac{\nabla^2\sqrt{P}}{\sqrt{P}} =0 \text{ or equivalently } \partial_t S + \frac{(\nabla S)^2}{2m}+V +Q =0
	\end{equation}
\end{subequations}

\noindent If we combine $P$ and $S$ into one complex field $\Psi$ by choosing
\begin{equation}
	\Psi(\mathbf r,t)\;=\;\sqrt{P(\mathbf r,t)}\,\exp\!\big(iS(\mathbf r,t)/\mu\big)
\end{equation}

\noindent and substitute into the $\mathcal{A}[P,S]$ we obtain the action of the Lagrangian in terms of $\Psi$ and $\Psi^*$.

\begin{equation}
	\label{eq:comL}
	\mathcal{A}[\Psi,\Psi^*] \;=\; \int dt\!\int d^3r\; \left\{
	\frac{i\mu}{2}\big(\Psi^*\partial_t\Psi - \Psi\partial_t\Psi^*\big)
	- \frac{\mu^2}{2m}|\nabla\Psi|^2 - V|\Psi|^2
	\right\}
\end{equation}
\noindent   The same extremisation procedure gives us the equation for $\Psi$
\begin{equation}
	\label{eq:genschro}
	\boxed{	i\mu\,\partial_t\Psi = -\frac{\mu^2}{2m}\nabla^2\Psi + V\Psi } 
\end{equation}

\noindent Here,  $\mu$ is an unknown constant. We observe that, if $\mu=0$, we get equations of classical mechanics, if $\mu=\hbar$ we get the well known Schr\"{o}dinger equation.
\vspace{0.5pc}

\section{Regularisation, a requirement in Bohmian trajectory description}
\label{sec3}

The Fisher information–augmented action introduced in Section~2.2 naturally
defines a \emph{density-based} variational problem, in which the probability
density $P$ (or equivalently $X=\sqrt{P}$) is treated as a fundamental field
variable. In this formulation, the functional is invariant under global
rescalings of the amplitude, $X \to \lambda X$ with constant $\lambda$,
so that its EL equation is insensitive to the overall
normalisation of $X$. Consequently, a normalisation constraint
$\int X^2\,dx=1$ must be imposed to fix the global scale of the solution.
When supplemented by this constraint, the EL equations acquire a
Sturm--Liouville structure with a spectral parameter emerging as a Lagrange
multiplier, leading to an equation of Ermakov--Pinney type for $X$ (\cite{Pinney1950, Reid1971}).
\vspace{0.5pc}

\paragraph{Ermakov--Pinney reduction from global amplitude admissibility.}
The regularisation requirement may be formulated directly at the field level by
treating the amplitude $X=\sqrt{P}$ as the fundamental dynamical variable.
Consider the global amplitude functional \ref{eq:density_action} with normalisation constraint
\begin{equation}
	\label{eq:1D_Field_constraint}
	\mathcal A_{\rm EP}[X;E]
	=
	\int X^2\left(
	\frac{\mu^2}{2m}\frac{X'^2}{X^2}
	+
	\frac{C^2}{2m}\frac{1}{X^4}
	+ V \right)dx -E\left(\int X^2dx-1\right)
\end{equation}

\noindent where the inverse-square term encodes the conserved stationary flux, and the
normalisation constraint is implemented through a Lagrange multiplier $E$ 
which subsequently plays the role of the spectral (energy) parameter in the
resulting EL equation.
\[
\frac{d}{dx}\left(\frac{\partial L}{\partial X'}\right)
-
\frac{\partial L}{\partial X}=0
\]
yields
\[
\boxed{X'' +\frac{2m}{\mu^2}(E-V(x))X=-\frac{C^2}{\mu^2}\frac{1}{X^3}}
\]

\[
\text{or} \quad \boxed{%
	X'' + \omega^2(x)\,X
	= -\frac{C^2}{\mu^2}\,\frac{1}{X^3}}
  \qquad
   \text{with}\qquad
    \omega^2(x):=\frac{2m}{\mu^2}\Bigl(E-V(x)\Bigr).
\]
which is the Ermakov--Pinney equation governing admissible amplitude
configurations. In the asymptotic regime where the potential term is locally
subdominant, this reduces to
\[
X''\simeq -\frac{C^2}{\mu^2}\frac{1}{X^3},
\]
whose regular branch produces the canonical scaling behaviour of the amplitude
near its nodes.
\vspace{0.5pc}

\noindent From the Bohmian viewpoint, the global field-theoretic reduction
corresponds to a deterministic trajectory regime in which the local momentum
field obeys the Hamilton--Jacobi energy balance, while the density
$P=X^2$ acts only as a normalisation field. The stationary flux closure
$P(x)p(x)=C$ then links the momentum and amplitude sectors pointwise,
so that admissible spatial variations of $p(x)$ must satisfy a local
regularisation principle compatible with the effective potential
$V_{\rm eff}=V+Q$. This principle is implemented below through a reduced
shell variational formulation with an effective Lagrangian of the form
$(T-V_{\rm eff})$.
\vspace{0.5pc}

\noindent It is therefore useful to distinguish two complementary aspects of the
variational description. The density-based Fisher information formulation yields a global
spectral problem with Sturm--Liouville structure and Ermakov--Lewis
invariants \cite{AAKErmakov}, while the flux-closed stationary reduction leads to a local
mechanical description in terms of spatial momentum flows, isolating the
regularisation mechanism and enabling explicit analytical solutions.
\vspace{0.5pc}

\paragraph{Hamilton--Jacobi (shell) momentum closure.}
Within the stationary Bohm--Madelung decomposition, the local momentum field
\(
p(x):=\partial_x S
\)
is a real kinematic quantity. The real part of the stationary equation takes
the Hamilton--Jacobi form of an energy balance,
\begin{equation}
	\label{eq:HJclosure}
	\frac{p^2(x)}{2m}+V_{\mathrm{eff}}(x)=E,
	\qquad 
	V_{\mathrm{eff}}(x):=V(x)+Q(x),
	\qquad 
	Q(x):=-\frac{\mu^2}{2m}\frac{X''(x)}{X(x)}.
\end{equation}
Thus, once the amplitude sector fixes the quantum potential \(Q[X]\), the momentum sector is closed pointwise by
\begin{equation}
	\label{eq:pfromVeff}
	p(x)=\pm\sqrt{2m\big(E-V_{\mathrm{eff}}(x)\big)}\,,
\end{equation}
on intervals where the radicand is nonnegative. This HJ closure provides the natural entry point for a reduced mechanical (shell) variational principle for the momentum field.
\vspace{0.5pc}

\paragraph{Reduced kinematic variational principle.}
To place the momentum closure \eqref{eq:HJclosure}--\eqref{eq:pfromVeff} in a
variational framework, we introduce a reduced functional
\(\mathcal A_{\rm red}[p]=\int dx\,L(p,p';x)\) or equivalently  \(\mathcal A_{\rm red}[X]=\int dx\,L(X,X';x)\) together with the probability
normalisation $\int P\,dx=1$. This functional arises after sequential reduction,
using the continuity relation to eliminate the amplitude variable, and its
EL equation yields a closed differential equation for the auxiliary momentum
field \(p(x)\). Upon substitution of the quantum potential \(Q[X]\), this
description is fully equivalent to the HJ energy balance in the
effective potential \(V_{\mathrm{eff}}=V+Q[X]\), and therefore provides a
mechanical encoding of the amplitude dynamics.
\vspace{0.5pc}

\noindent The use of a reduced (shell) variational principle has a direct
analogue in classical mechanics: Jacobi’s principle provides a fixed-energy
configuration-space formulation in which the momentum field is determined
without explicit time parametrisation. In the Bohm--Madelung setting, the
stationary real equation assumes this HJ form with an effective
potential including the quantum term. The reduced functional introduced here
should therefore be regarded as a mechanical encoding of the HJ closure,
rather than an additional assumption.
\subsection{Preliminary Variational Structure and Equivalent Forms}
\label{subsec:prelim_variational}
\vspace{0.5pc}

Before introducing the constrained variational formulation,
we briefly clarify the structure of the amplitude term
appearing in the quantum potential.
\vspace{0.5pc}

\noindent The second-order quantity
\begin{equation}
	\frac{X''}{X}
\end{equation}
admits the identity
\begin{equation}
	\frac{X''}{X}
	=
	\left(\frac{X'}{X}\right)'
	+
	\left(\frac{X'}{X}\right)^2.
	\label{eq:log_identity}
\end{equation}
The first term on the right-hand side is a total derivative.
As discussed in standard variational treatments \cite{GoldsteinCM},
adding a total derivative $dF/dx$ to a Lagrangian
does not alter the EL equations.
Consequently, for variational purposes,
the second-order expression $X''/X$
is equivalent to retaining only the quadratic term
\begin{equation}
	\left(\frac{X'}{X}\right)^2 .
\end{equation}

\noindent This justifies using a Lagrangian depending only on
$X$ and $X'$ without explicit second derivatives.

\paragraph{Equivalent variational reductions.}
The constrained minimisation may be carried out either in terms of
\begin{equation}
	L(X,X';x)
	\quad\text{or}\quad
	L(p,p';x),
\end{equation}
since the stationary flux condition
\begin{equation}
	p = \frac{C}{X^2}
\end{equation}
provides a one-to-one transformation between the two descriptions.
In both formulations, the EL reduction
leads to the same first integral.  
\vspace{0.5pc}

\noindent It should be noted that the amplitude equation follows directly from the
variational principle associated with the Fisher information action. This
structure naturally enforces amplitude admissibility through regularity and
normalisation, leading to a globally well-posed spectral problem.
\vspace{0.5pc}

\subsection{Constraint Variational Regularisation and Canonical Shell Branch}
\label{subsec:constraint_regularisation}

We introduce a constrained variational formulation based solely on amplitude
admissibility. In this construction, current conservation is not imposed as an
external constraint; instead, the admissible branches are determined purely by
normalisation and regularity principles.
\vspace{0.5pc}

\noindent  Consider a one--dimensional stationary ``amplitude'' field $X(x)>0$,
together with a conserved stationary flux encoded through a momentum field
\begin{equation}
	p(x)=\frac{C}{X(x)^2},
	\qquad C\in\mathbb{R}\setminus\{0\},
	\label{eq:pC_over_X2}
\end{equation}
which is the standard Madelung/Bohm continuity reduction for a static current.
We introduce the constrained action
\begin{equation}
	\mathcal A[X;E]
	=
	\int dx\,
	\left[
	\frac{p^2}{2m}
	+
	\frac{\mu^2}{2m}\left(\frac{X'}{X}\right)^2
	-
	V(x)
	\right]
	-
	E\left(\int X^2\,dx-1\right),
	\label{eq:sigma_action_correct}
\end{equation}
where $m>0$ is the mass parameter, $\mu$ is the Fisher information coupled quantum potential parameter,
and $E$ is a Lagrange multiplier enforcing normalisation $\int X^2 dx=1$.
Inserting \eqref{eq:pC_over_X2} into \eqref{eq:sigma_action_correct} yields
\begin{equation}
	\mathcal A[X;E]
	=
	\int dx\,
	\left[
	\frac{C^2}{2m\,X^4}
	+
	\frac{\mu^2}{2m}\frac{X'^2}{X^2}
	-
	V(x)
	-
	E X^2
	\right]
	\;+\;E.
	\label{eq:sigma_action_reduced}
\end{equation}
The shell-level Lagrangian uses the mechanical form $L=T-V_{\rm eff}$ together
with the normalisation constraint, so that the local dynamics of $p(x)$ are
determined through their coupling to the amplitude field $X$.
\paragraph{Euler--Lagrange equation.}
The reduced Lagrangian density is (without an additive constant $E$)
\begin{equation}
	L(X,X';x)
	=
	\frac{C^2}{2m\,X^4}
	+
	\frac{\mu^2}{2m}\frac{X'^2}{X^2}
	-
	V(x)
	-
	E X^2.
	\label{eq:L_reduced}
\end{equation}
Since $V(x)$ is independent of $X$, it plays no role in the EL equation. 
\vspace{0.5pc}

\noindent A direct computation gives
\begin{equation}
	\boxed{
		X''-\frac{X'^2}{X}
		+\frac{2C^2}{\mu^2}\frac{1}{X^3}
		+\frac{2mE}{\mu^2}X^3
		=0.
	}
	\label{eq:EL_X}
\end{equation}
Variation with respect to $E$ yields the normalisation constraint
\begin{equation}
	\int X^2\,dx=1.
	\label{eq:constraint_norm}
\end{equation}

\paragraph{Quadrature from the EL equation.}
Introduce $u:=X'/X$. Then $X''-(X'^2/X)=u'X$, and
\eqref{eq:EL_X} reduces to
\begin{equation}
	u'
	+
	\frac{2C^2}{\mu^2}X^{-4}
	+
	\frac{2mE}{\mu^2}X^{2}
	=0.
	\label{eq:uprime}
\end{equation}
Using $u'=\frac{du}{dX}(uX)$ and integrating, we obtain the
first-order quadrature
or equivalently,
\begin{equation}
	\left(\frac{X'}{X}\right)^2
	=
	\frac{C^2}{\mu^2}\frac{1}{X^{4}}
	-\frac{2mE}{\mu^2}X^{2}
	+2\kappa,
	\qquad \kappa\in\mathbb{R}.
	\label{eq:first_integral_u}
\end{equation}
\begin{equation}
	\boxed{
		X'^2
		=
		\frac{C^2}{\mu^2}\frac{1}{X^{2}}
		-\frac{2mE}{\mu^2}X^{4}
		+2\kappa X^{2}.
	}
	\label{eq:first_integral_X}
\end{equation}

\noindent In the present subsection we focus on the generalised regularisation branch $\kappa \neq 0$,
\vspace{0.5pc}

\noindent for which \eqref{eq:first_integral_X} becomes a cubic (elliptic) quadrature.

\begin{equation}
	\left(\frac{X'}{X}\right)^2
	=
	\frac{C^2}{\mu^2}\frac{1}{X^{4}}
	-\frac{2mE}{\mu^2}X^{2}
	+2\kappa,
	\label{eq:first_integral_u_correct}
\end{equation}

\begin{equation}
	X'^2
	=
	\frac{C^2}{\mu^2}\frac{1}{X^{2}}
	-\frac{2mE}{\mu^2}X^{4}
	+2\kappa X^{2},
	\label{eq:first_integral_X_correct}
\end{equation}

\begin{equation}
	P'^2
	=
	\frac{4C^2}{\mu^2}
	+8\kappa P^{2}
	-\frac{8mE}{\mu^2}P^{3},
	\qquad
	P:=X^{2}.
	\label{eq:cubic_P_correct}
\end{equation}

\subsection{Elliptic form}
\label{subsec:elliptic_kappa}

Set $P:=X^2>0$ represents the probability density (up to normalisation). Then $X'^2=\dfrac{P'^2}{4P}$ and
\eqref{eq:first_integral_X} with $\kappa \neq 0$ becomes
\begin{equation}
	\boxed{
		P'^2
		=
		4\left(A+2\kappa\,P^2
		-
		B\,P^3\right),
		\qquad
		A:=\frac{C^2}{\mu^2},\quad
		B:=\frac{2mE}{\mu^2}>0.
	}
	\label{eq:Pprime_kappa_correct}
\end{equation}
This is a cubic (elliptic) differential equation in $P(x)$.

\paragraph{Reduction to Weierstrass normal form.}
Equation \eqref{eq:Pprime_kappa_correct} can be reduced to canonical Weierstrass form through successive transformation from a general equation given by

\begin{equation}
	P'^2=aP^3+bP^2+cP+d , \quad \text{ with } a = -\frac{8mE}{\mu^2}, \quad b = 8\kappa, c = 0, \text{ and } d=\frac{4C^2}{\mu^2}
\end{equation}
\[
P = u -\frac{b}{3a} = u+\delta, \text{ where } \delta =\frac{\kappa\mu^2}{3mE}
\]
shifts converts to depressed-cubic condition ($e_1+e_2+e_3=0$). Making another scaling transformation $\xi = \sqrt{\frac{2mE}{mu^2}}(x-x_0)$, with $x_0$ as an arbitrary constant, we can reduce to $P(x)=u(\xi)+\delta$

Setting
\begin{equation*}
	P(x)= u(\xi) +\delta 
	\text{ where } u(\xi) \quad \text { satisfies } \quad
	\left(\frac{du}{d\xi}\right)^2 = 4u^3(\xi)-g_2u(\xi)-g_3,
	\label{eq:wp_solution_kappa_correct}
\end{equation*}

\noindent one verifies directly that \eqref{eq:Pprime_kappa_correct} is satisfied.
The invariants are therefore

\begin{equation}
	g_2 = \frac{4}{3}\frac{\kappa^2 \mu^4}{m^2E^2}, \qquad
	g_3 = -\left(\frac{2C^2}{mE}+\frac{8}{27}\frac{\kappa^3\mu^6}{m^3E^3}\right)
	\label{eq:g2g3_kappa0}
\end{equation}

\noindent Here $u(\xi)=\wp(\xi;g_2,g_3)$ denotes the Weierstrass elliptic function \cite{handbook} and $x_0$ is an
integration constant fixed by boundary conditions.

\paragraph{Local canonical (regular) branch.}
The admissible regular branch corresponds to $X(x)\to0$ at some finite
point $x\to x_0^+$, hence $P(x)\to0$.  From
\eqref{eq:cubic_P_correct},
\begin{equation}
	P'^2 \;\xrightarrow[P\to0]{}\; 4A,
	\qquad\Rightarrow\qquad
	P'(x)\to 2\sqrt{A}=\frac{2|C|}{\mu}.
\end{equation}
Therefore the local expansion is
\begin{equation}
	P(x)=X(x)^2 \sim \frac{2C}{\mu}(x-x_0),
	\qquad (X\to0).
	\label{eq:P_linear_asympt}
\end{equation}
This implies the universal square--root behavior
\begin{equation}
	X(x)\sim
	\sqrt{\frac{2|C|}{\mu}}\;\sqrt{x-x_0}.
	\label{eq:X_sqrt_asympt_correct}
\end{equation}

\paragraph{Canonical momentum relation.}
Using the conserved flux relation $p=C/X^2$, the asymptotics
\eqref{eq:P_linear_asympt} give
\begin{equation}
	p(x)
	=
	\frac{C}{X(x)^2}
	\sim
	\frac{C}{\frac{2C}{\mu}(x-x_0)}
	=
	\frac{\mu}{2(x-x_0)}.
\end{equation}
Hence the canonical product approaches the universal constant
\begin{equation}
	\boxed{
		p(x)\,(x-x_0)\;\longrightarrow\;\frac{\mu}{2}
		\qquad (x\to x_0^+).
	}
	\label{eq:canonical_px_correct}
\end{equation}
Choosing $x_0=0$ (or absorbing the shift into the origin) gives the canonical
regularisation condition
\begin{equation}
	p(x)\,x=\frac{\mu}{2},
	\label{eq:canonical_relation_final_correct}
\end{equation}
which emerges purely from the constrained variational principle
\eqref{eq:sigma_action_correct} together with the stationary flux reduction
\eqref{eq:pC_over_X2}.
\vspace{0.5pc}

\noindent For $\kappa\neq 0$, the quadrature involves the full cubic in $P=X^2$, leading to a shifted Weierstrass form with three non degenerate roots. These branches represent a broader class of canonical regularisation solutions whose global geometry is governed by the root structure of the cubic, while the local asymptotic regularisation remains unchanged.

\paragraph{Interpretation.}
The canonical relation \eqref{eq:canonical_relation_final_correct}
is thus not imposed ad hoc,
but emerges as the admissible asymptotic solution
of the constrained variational principle
\eqref{eq:sigma_action_correct}.
The general solution of \eqref{eq:first_integral_X}
is elliptic in nature (for $\kappa\neq0$),
while the canonical branch corresponds to the degenerate sector
$\kappa=0$.
Importantly, the local behaviour \eqref{eq:X_sqrt_asympt_correct}
and hence \eqref{eq:canonical_relation_final_correct}
remain valid even for non-constant potentials,
since the inverse-power term dominates near $X\to0$.
\vspace{0.5pc}

\paragraph{Classical limit \(\mu\to 0\) from the variational principle.}
The regularisation enters the theory exclusively through the Fisher information term
\(\frac{\mu^2}{2m}\left(\frac{X'}{X}\right)^2\) in the action functional
\eqref{eq:sigma_action_correct}. In the limit \(\mu\to 0\), this gradient term vanishes
identically and the functional loses all dependence on \(X'\). The EL
equation with respect to \(X\) then becomes purely algebraic rather than differential,
implying that the amplitude \(X(x)\) is no longer dynamically constrained by a
regularity condition. Consequently, the remaining dynamics is governed solely by the
classical Hamilton--Jacobi relation encoded in \(p(x)\). Thus the regularised
variational problem collapses smoothly to classical mechanics in the strict
\(\mu=0\) limit, demonstrating that regularisation is neither required nor operative
in the classical regime.

\paragraph{Ontological remark.}
In the Bohmian framework adopted here,
the shell Hamiltonian generates deterministic trajectories
independently of any normalisation requirement.
However, deterministic evolution alone does not guarantee
admissibility of amplitude configurations.
The constrained variational formulation introduced above
acts as a guiding principle,
selecting physically admissible branches of motion.
In particular, the canonical relation
\(
p\,x=\mu/2
\)
emerges as a consequence of amplitude regularisation,
rather than as an external postulate.
Thus, the normalisation constraint is not merely technical but ontological:
it renders the Bohmian amplitude dynamics well-posed, selecting admissible
branches of the scale-invariant quantum potential and enabling analytical
reductions that would otherwise remain underdetermined.

\section{Regularised dBB wave function for standard potentials}
\label{sec4}
\noindent In this section we illustrate the regularised stationary solutions using the one-dimensional harmonic oscillator as a representative example. The same variational and shell-reduction principles extend to other potentials and to separable components in higher dimensions. Table~\ref{tab1} summarises common potentials that admit regularisation within the stationary de Broglie--Bohm description, while Table ~\ref{tab:spectrum_comparison} compares the resulting energy spectra with their standard quantum-mechanical counterparts, highlighting the associated spectral shifts.

\subsection{The Harmonic Oscillator}
\vspace{0.5pc}

The specific Hamiltonian of the one-dimensional harmonic oscillator is 
\begin{subequations}
	\begin{equation}
		\frac{\mu^2}{8mx^2}+\frac{1}{2}k_sx^2-\frac{\mu^2}{2m}\frac{X''(x)}{X(x)}=E
	\end{equation}
	where $k_s$ is the spring constant, and the first term arises from the canonical regularisation condition $p(x)x=\mu/2$. With some rearrangement of the terms, we get
	\begin{equation}
		X''(x) - X(x) \left( \frac{1}{4x^2} + \kappa x^2 - k^2 \right) = 0
	\end{equation}	
\end{subequations}
\noindent where $\kappa = \frac{mk_s}{\mu^2}$ and $k^2=\frac{2mE}{\mu^2}$.
\vspace{0.5pc}

\noindent The differential equation above can be solved by choosing an appropriate substitution for $X(x)$ given by 
\begin{subequations}
	\begin{equation}
		X(x) = x^{1/2} e^{-\frac{\sqrt{\kappa}}{2} x^2} u(x)
	\end{equation}
	\noindent This simplifies to Hermite-like differential equation for $u(x)$:
	\begin{equation}
		u''(x) - 2 \sqrt{\kappa} x u'(x) + \lambda u(x) = 0
	\end{equation}
	\noindent for some constant $\lambda$ depending on $k$ and $\kappa$.
	\vspace{0.5pc}
	\noindent To bring it exactly into Hermite form, one can rescale by choosing a new transformation given by
	$z = \sqrt[4]{\kappa} x$, which results in	
	\begin{equation}
		\frac{d^2 u}{dz^2} - 2z \frac{du}{dz} + \left( \frac{k^2}{\sqrt{\kappa}} - 1 \right) u = 0
	\end{equation}
\end{subequations}	
\noindent The above equation matches Hermite differential equation \cite{handbook}, if we define:
\begin{equation}
	\frac{k^2}{\sqrt{\kappa}} - 1 = 2n \quad \Rightarrow \quad k^2 = \sqrt{\kappa}(2n + 1),\quad n = 0, 1, 2, \dots
\end{equation}
\noindent Thus, the energy of the harmonic oscillator is
\begin{equation}
	\label{eq:modHME}
	\boxed{E_n = \mu \omega \left(n + \frac{1}{2}\right)} \quad,  \text{   with  } \omega = \sqrt{\frac{k_s}{m}}
\end{equation}

\noindent and wave function obtained from the generalised EL equation of harmonic potential 
\begin{equation}
	\label{eq:modHwf}
	{X(x) = X_0x^{1/2} e^{-\frac{\sqrt{\kappa}}{2} x^2} H_n(\sqrt[4]{\kappa} x)}
\end{equation}
where $X_0$ is the normalisation constant.
\newpage

\vspace{.5pc}
\begin{table}[!h]
	\begin{threeparttable}
	\caption{Summary of regularised solution for various potentials}
	\centering
	\begin{tabularx}{\textwidth}{l l X l}
		\hline
		Potential & Parameters & Amplitude function $\psi_i(q_i)^\dagger$ & Spectra \\
		\hline
		Free Particle  & $k=\sqrt{\frac{2mE}{\mu^2}}$ &
		$\sqrt{x}\,J_{\frac{1}{\sqrt{2}}}(kx)^\ddagger$ & $E$ \\
		
		Har.\ Osc.\ $(\tfrac{1}{2}k_sx^2)$ &
		$\kappa = \frac{mk_s}{\mu^2}$ &
		$ x^{1/2} e^{-\frac{\sqrt{\kappa}}{2} x^2}
		H_n(\sqrt[4]{\kappa} x)^\ddagger$ &
		$E_n={\mu \omega}\!\left(n+\tfrac{1}{2}\right)$ \\
		
		Coul.\ Pot.\ $(-\alpha/r)$ &
		$\kappa = \sqrt{-\frac{2mE}{\mu^2}} > 0$,
		$\kappa_w = \frac{m\alpha}{\mu^2\kappa}$ &
		$\sqrt{x}\,M_{\kappa_w,\,1/\sqrt{2}}(2\kappa x)^\ddagger$ &
		$E_n = - \frac{m\alpha^2}{2\mu^2
			\left(n+\frac{1}{\sqrt{2}}+\tfrac{1}{2}\right)^2}$ \\
		
		Rad.\ Har.\ Osc.\ (2D) &
		$\gamma = \frac{m\omega}{\mu},\quad
		\omega = \sqrt{\frac{\kappa}{m}}$ &
		$ r^{l+\frac{1}{2}}e^{-\frac{1}{2}\gamma r^2}
		L_n^{(l)}(\gamma r^2)^\ddagger$ &
		$E_n={\mu\omega}(2n+l+1)$ \\
		
		3D Rad.\ Coul.\ Pot.\ $(-\alpha/r)$ &
		$\mu_w =\sqrt{l(l+1)+\tfrac12}$ &
		$ \frac{1}{r}
		M_{\kappa_w,\mu_w}(2\kappa r)$ &
		$E_{n,l}=-\frac{m\alpha^2}{2\mu^2
			\left(n+\mu_w+\tfrac12\right)^2}$ \\
		
		3D Ang.\ Coul.\ Pot. &
		$m_\phi$ = Az.\ param. &
		$ P_l^{\,m_\phi+\frac12}(\cos\theta)^\ddagger$ &
		Not Applicable \\
		\hline
	\end{tabularx}
	\label{tab1}
	\begin{tablenotes}[flushleft]
	\item {\small $^\dagger$ All amplitudes are given up to an overall normalisation constant. $^\ddagger$ References to special functions used above---see~\cite{Arfken}. }
	\end{tablenotes}
	\end{threeparttable}
\end{table}
\noindent Table~\ref{tab:spectrum_comparison} shows that the shell-regularised formulation
preserves the standard spectra for systems without singular radial behaviour,
such as constant potentials and harmonic oscillators (up to two dimensions), while introducing
systematic shifts in Coulomb-type problems where inverse-square structures
play a dominant role. These shifts originate from the regularising Fisher term,
which effectively modifies the centrifugal contribution and removes nodal
singularities, leading to analytically solvable spectra that remain structurally
close to their quantum-mechanical counterparts.

\begin{table}[!h]
	\centering
	\begin{threeparttable}
	\caption{Comparison of energy spectra between standard quantum mechanics (QM) and
		shell-regularised Bohmian formulation for representative solvable systems.
		Here $\Delta E := E_{\text{Shell}}-E_{\text{QM}}$.}
	\begin{tabularx}{\textwidth}{l l X c}
		\hline
		System & $E_{\text{QM}}$ & $E_{\text{Shell}}$ & $\Delta E$ \\
		\hline
		Constant Potential  & $E - V_0$ & $E - V_0$ & $0$ \\
		
		Har. Osc. (1D) 
		& $\hbar\omega\left(n+\tfrac{1}{2}\right)$ 
		& $\hbar\omega\left(n+\tfrac{1}{2}\right)$ 
		& $0$ \\
		
		Har. Osc. (2D) 
		& $\hbar\omega\left(2n_r+|m|+1\right)$ 
		& $\hbar\omega\left(2n_r+l+1\right)$ 
		& $\approx 0$ (notation shift) \\
		
		Har. Osc. (3D) 
		& $\hbar\omega\left(2n_r+l+\tfrac{3}{2}\right)$ 
		& $\hbar\omega\left(2n_r+1+\sqrt{l(l+1)+\tfrac{1}{2}}\right)$ 
		& $\neq 0$ \\
		
		Coul.\ Pot. (1D) 
		& $-\dfrac{m\alpha^2}{2\hbar^2 n^2}$ 
		& $-\dfrac{m\alpha^2}{2\hbar^2\left(n+\tfrac{1}{\sqrt{2}}+\tfrac{1}{2}\right)^2}$ 
		& $\neq 0$ \\
		
		Coul.\ Pot. (2D) 
		& $-\dfrac{m\alpha^2}{2\hbar^2\left(n_r+|m|+\tfrac{1}{2}\right)^2}$ 
		& $-\dfrac{m\alpha^2}{2\hbar^2\left(n_r+l+1\right)^2}$ 
		& relabelling and effective $+\tfrac{1}{2}$ shift \\
		
		Coul.\ Pot. (3D)$^*$ 
		& $-\dfrac{m\alpha^2}{2\hbar^2\left(n_r+l+1\right)^2}$ 
		& $-\dfrac{m\alpha^2}{2\hbar^2\left(n_r+\mu_w+\tfrac{1}{2}\right)^2}$ 
		& $\neq 0$ \\
		\hline
	\end{tabularx}
	\label{tab:spectrum_comparison}
	\begin{tablenotes}[flushleft]
		\item {\small $^*$ $\mu_w =\sqrt{l(l+1)+\tfrac{1}{2}}$, obtained from Whittaker convergence condition.
		Here $n$ denotes the principal quantum number in one dimension, while $n_r$ represents the radial quantum number in higher-dimensional separable systems and energy parameters are switched to $\hbar$ units.}
	\end{tablenotes}
	\end{threeparttable}
\end{table}

\begin{figure}[h!]
	\centering
	\includegraphics[width=\linewidth]{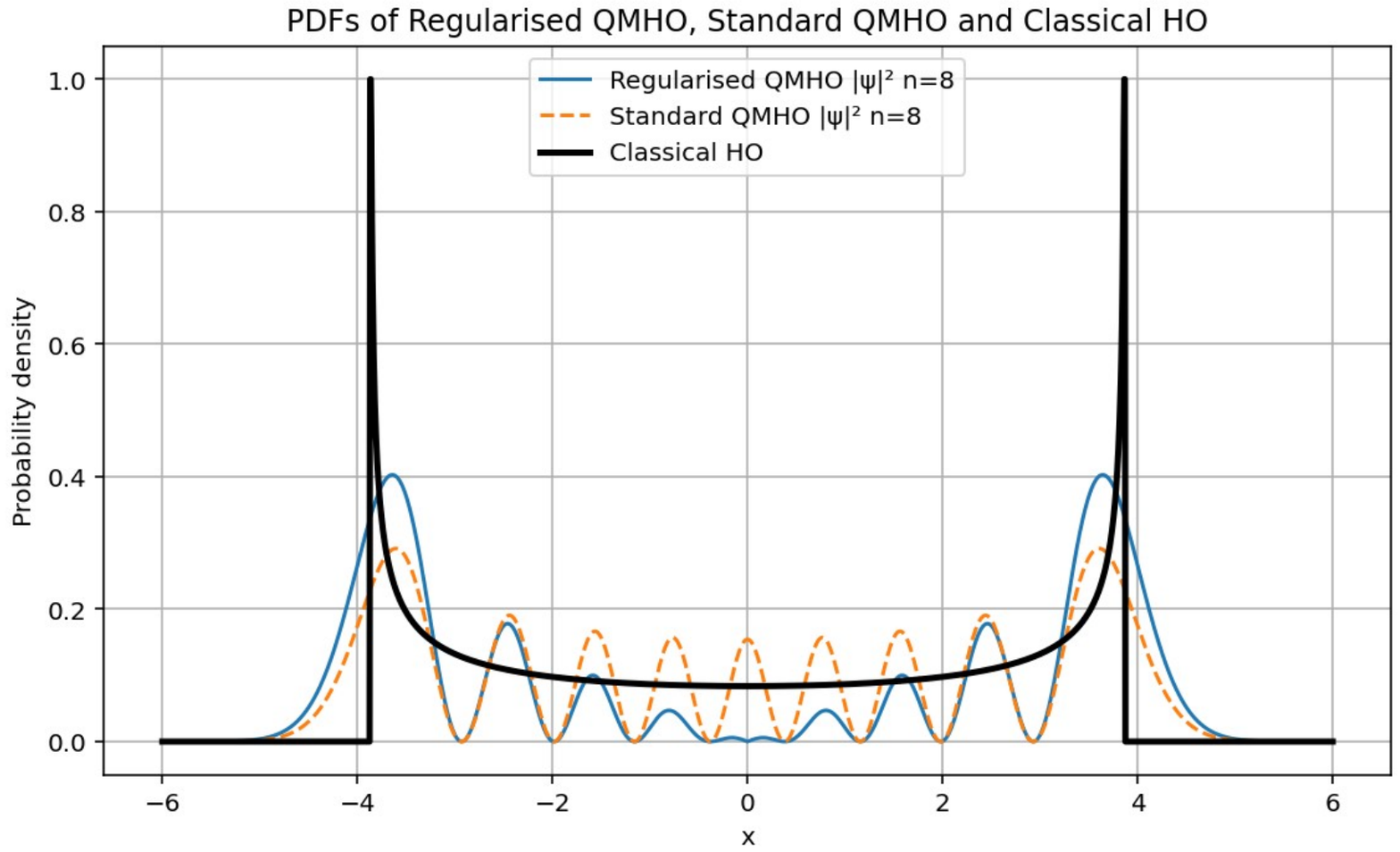}
	\caption{Regularised and standard distributions for $n=8$ over classical envelope.}
	\label{Fig1}
\end{figure}

\noindent A typical comparison of the probability distribution function for the
harmonic oscillator for classical, traditional quantum mechanics and
regularised Bohmian mechanics is shown in Fig. (\ref{Fig1}). There are subtleties in the way
the amplitudes are regulated in Bohmian mechanics, which manifest notably at
the origin through a vanishing probability density at $x=0$ and evenness of the probability density for all values of 
$n$. This behaviour arises from the
inverse-square regularising term, which enforces a finite and admissible
stationary current and selects the canonical regular branch of the solution.
The boundaries $x=\pm a$ follow a similar asymptotic behaviour to that obtained
in standard quantum mechanics.
\vspace{0.5pc}

\noindent From an analytical standpoint, the amplitude function $X(x)$ obtained through
the regularised EL equations is naturally defined on a single
analytic branch (for example $x>0$) where it remains positive and normalisable.
The extension to negative values of $x$ is achieved by reconstructing the full
wave function $\psi(x)=X(x)e^{iS(x)/\mu}$, whereby the phase absorbs the sign
change of the coordinate without affecting physical observables such as
$\nabla S$ or the probability density. Thus, the regularised harmonic-oscillator
solutions should be interpreted as branch-defined analytic amplitudes whose
global behaviour over $x\in\mathbb{R}$ is recovered through analytic continuation
of the phase factor. This branch structure is a direct consequence of the
canonical regularisation condition $p(x)x=\mu/2$ and is consistent with the
invariant-based reduction developed in the preceding sections.

\section{Remarks on the regularised wave function analytical branches}
\label{sec5}

The wave functions obtained from the variational minimisation possess a
`square-root' prefactor, restricting the amplitude to a single analytic
branch (e.g.\ $q>0$). Extension to negative $q$ is achieved by reconstructing
the full wave function $\psi=Xe^{iS/\mu}$, whereby the sign change of the
square-root amplitude is absorbed into a constant phase factor. This
continuation does not affect physical observables such as $\nabla S$ or the
probability density. The wave function therefore exhibits a natural branch
point at $q=0$, remaining continuous but non-differentiable.
\vspace{0.5pc}

\noindent The canonical regularisation $p(q)\,q=\mu/2$ implies $p(q)\simeq \mu/(2q)$ near a
node, so that the phase satisfies $S(q)=\int p\,dq \simeq (\mu/2)\ln|q|$ on each
analytic branch. Extending the wave function to $q<0$ may then be implemented by
a branch-continuation phase jump $S\to S\pm \mu\pi/2$, i.e.\ $\psi\to (\pm i)\psi$,
which leaves $|\psi|^2$ and observables depending on $\nabla S$ unchanged.
\vspace{0.5pc}

\noindent Further, the phase part retains the classical action–angle contribution together with
an additional constant phase arising from the canonical regularisation
condition $pq=\mu/2$, entering the exponential factor as $e^{ipq/\mu}$.
Depending on the potential and the corresponding analytic solution, this
phase is further tied to the index of the associated special function through
analytic branch continuity; for example, a Bessel-type solution produces a
shift of order $\pi\nu$, while a harmonic-oscillator solution yields a shift
of order $\pi n$.
\vspace{0.5pc}

\noindent The energy spectra obtained for various potentials show slight shifts in the values as shown in the ``Spectra--Column" of Table \ref{tab1}. However, it does not alter the form of the spectra (as shown in sections [\ref{sec3}]). It  is attributable to the term $1/q^2$ in the respective differential equations obtained from minimisation which acts like a centrifugal barrier type of potential that is found in 2D and 3D polar coordinate systems and thus \textit{regularises} the wave functions. Parenthetically, we note that the nature of inverse square potential has been studied as a pathological case  to address some of the important aspects of renormalisation, regularisation, symmetry breaking and conformal invariance \cite{essin}. 
\vspace{0.5pc}

\noindent Further, we observe that the traditional method of solving the system of guiding and continuity equations through a numerical procedure would recover only approximate PDFs, which are effectively envelopes of PDFs obtained analytically in the previous sections.

\section{Compton length}
\label{sec6}

\subsection*{Discriminant scale and emergence of the Compton length}

For the general $\kappa\neq 0$ branch, the first integral, (\ref{eq:first_integral_X}) in terms of
$P=X^2$ (probability density) takes the cubic form
\begin{equation}
	P'^2 = 4A + 8\kappa P^2 - 4B P^3,
	\qquad
	A=\frac{C^2}{\mu^2},\quad
	B=\frac{2mE}{\mu^2}.
	\label{eq:cubic_general}
\end{equation}
The associated elliptic curve is governed by the cubic polynomial
\(
\mathcal{P}(P)=-B P^3 + 2\kappa P^2 + A. 
\)
Its discriminant is
\begin{equation}
	\Delta = -A\left(32\kappa^3 + 27A B^2\right).
	\label{eq:discriminant}
\end{equation}
Degeneracy of the elliptic curve (coalescing roots) occurs when
\(
\Delta=0
\),
that is,
\begin{equation}
	32\kappa^3 + 27A B^2 = 0.
	\label{eq:degeneracy_condition}
\end{equation}
Solving for $\kappa$ gives the characteristic geometric scale
\begin{equation}
	|\kappa| \sim \left(\frac{27}{32}A B^2\right)^{1/3}.
\end{equation}
Since $\kappa$ carries dimensions of inverse length squared, one obtains
a critical length scale
\begin{equation}
	L_{\mathrm{cr}}
	\sim
	|\kappa|^{-1/2}
	\sim
	\left(A B^2\right)^{-1/6}.
	\label{eq:Lcr_general}
\end{equation}
Substituting $A=C^2/\mu^2$ and $B=2mE/\mu^2$ yields
\begin{equation}
	L_{\mathrm{cr}}
	\sim
	\frac{\mu}{(mE)^{1/3} C^{1/3}}.
\end{equation}
The physical identification $\mu=\hbar$,
$E\sim mc^2$, and $C/m\sim c$ leads to the emergence of the reduced
Compton wavelength as the characteristic geometric scale,

\begin{equation}
	\boxed{
		L_{\mathrm{cr}} \sim \frac{\hbar}{mc}
	}
\end{equation}
Thus, the degeneracy threshold of the elliptic reduction yields a characteristic
length scale of order the reduced Compton wavelength. This scale marks the
geometric threshold at which the canonical regularisation branch emerges.
\vspace{0.5pc}

\noindent We emphasise that the appearance of the Compton scale here is suggestive: the present framework is non-relativistic, and the identification is a dimensional (or parametric) correspondence rather than a derivation of relativistic dynamics.

\subsection*{Foundational interpretation}

In the constrained density-based variational formulation, the regularity
mechanism is not imposed as an interpretive postulate about uncertainty,
but arises dynamically from the interplay between (i) a conserved stationary
flux and (ii) the Fisher information gradient penalty
$\propto \mu^2(\partial_x\ln X)^2$. The resulting Euler--Lagrange
structure leads to an elliptic quadrature whose admissible branch is
selected by root geometry and asymptotic regularity at $X\to0$. On this
canonical branch one obtains a universal local scaling
$X^2\sim (2C/\mu)(x-x_0)$ and hence a canonical product
$p(x)(x-x_0)\to \mu/2$. Interpreting $X^2$ as a normalised density ties
the integration constants to geometric length scales: the
discriminant-controlled transition between elliptic regimes yields a
critical length $L_{\mathrm{cr}}$ determined by $\mu$, the mass
parameter, and the flux scale. In the special identification
$\mu=\hbar$ and $C/m\sim c$, $E\sim mc^2$, one recovers
$L_{\mathrm{cr}}\sim \hbar/(mc)$, i.e. the Compton scale, as a natural
geometric barrier. This suggests a structural stability mechanism 
emerging from variational regularity and conserved flux—by which
short-distance limitations can be viewed as geometric/functional
constraints of the density dynamics, rather than appended interpretive
principles or ad hoc modifications of classical electrodynamics.

\section{Summary and Conclusions}

\noindent The primary objective of this work has been to demonstrate that a systematic
regularisation procedure makes it possible to obtain analytical solutions of
the stationary de~Broglie--Bohm  equations, for a restricted
class of boundary conditions admitting stationary states. The central idea is
that stationary Bohmian dynamics possess global symmetry constraints and
invariant structures that can be exploited to reduce the coupled nonlinear
Madelung equations to analytically solvable forms. The present regularisation
framework identifies and utilises these invariant structures to obtain
well-behaved admissible solutions consistent with deterministic Bohmian
trajectories and normalisable probability densities.
\vspace{0.5pc}

\noindent The formulation begins with a Fisher information–augmented variational action
for the probability density and phase fields, from which the Madelung
equations follow variationally. At a global level, admissible amplitude
configurations are characterised by an Ermakov--Pinney reduction, encoding
regularity and finite trajectory behaviour through an invariant structure.
This global amplitude admissibility is realised deterministically at the
trajectory level through a reduced shell variational principle, in which the
local momentum field satisfies a Hamilton--Jacobi closure in an effective
potential. The resulting constrained dynamics select canonical regularisation
branches and enable analytical solutions of the stationary de~Broglie--Bohm
equations for a broad class of potentials.
\vspace{0.5pc}

\noindent A key development of the present work is the introduction of a reduced (shell)
variational principle for the local momentum field obtained through stationary
flux closure. This reduction isolates the regularisation mechanism at the level
of spatial momentum flow and leads to constrained Euler--Lagrange equations that
govern admissible amplitude configurations. The resulting first integral admits
an elliptic (Weierstrass) structure, revealing a broad family of analytical
solution branches compatible with stationary invariants.
\vspace{0.5pc}

\noindent The admissible asymptotic branch near amplitude zeros yields a universal canonical
relation $p(x)\,x \to \mu/2$, which emerges dynamically from amplitude regularity
and conserved flux rather than being imposed externally. This canonical Bohmian
regularisation introduces an inverse-square contribution in the effective
potential, ensuring finite, well-behaved solutions and enabling closed-form
analytical wave functions for standard stationary potentials. While the resulting
spectra retain the same structural form as in standard quantum mechanics, the
functional form of the wave functions reflects the intrinsic regularisation
imposed by the invariant structure.
\vspace{0.5pc}

\noindent Furthermore, the elliptic reduction possesses a discriminant-controlled geometric
scale. Under the physical identification $\mu=\hbar$ and $E\sim mc^2$, this scale
reduces naturally to the reduced Compton wavelength, indicating that the
short-distance behaviour of stationary Bohmian amplitudes can be interpreted as
a consequence of the underlying variational invariants rather than as an imposed
interpretive constraint.
\vspace{0.5pc}

\noindent In summary, the present work shows that the canonical regularisation method
provides a principled route to analytical solutions of stationary dBB equations
by exploiting the global symmetry and invariant structure of stationary states.
The approach yields structurally stable and normalisable solutions, clarifies
the role of inverse-square regularising terms, and establishes a unified
variational framework connecting Fisher information, Hamilton--Jacobi closure,
and exact solvable branches of Bohmian dynamics.
\vspace{0.5pc}

\noindent Future work will extend this invariant-based regularisation scheme to more
general potentials, higher-dimensional separable systems, and numerical studies
of guidance and continuity equations to further assess its applicability beyond
the stationary regime.

\section*{Declaration of Interests}

The research is an independent idea of the principal author before his affiliation with IBM Research, USA. Further, the authors declare that they have no known competing financial interests or personal relationships that could have appeared to influence the work reported in this paper. 
\vspace{0.5pc}

\section*{Use of Generative AI Tools}

Generative artificial intelligence tools were used solely for assistance in
language editing, LaTeX formatting, and manuscript structuring. The scientific
ideas, analytical derivations, and physical interpretations presented in this
work were independently developed and validated by the authors.
\section*{Acknowledgements}

The authors thank Professors S.V. Bhat (INSA Scientist, Department of Physics, Indian Institute of Science, Bangalore, India) and Radhakrishna Srikanth (Associate Professor of Theoretical Sciences, Poornaprajna Institute of Scientific Research, Bangalore, India) for their valuable discussions and suggestions.

\end{document}